\documentclass{aa}
\usepackage{graphicx}
\usepackage{color}
\usepackage[varg]{txfonts}

\begin{document}

\title{Forming pressure-traps at the snow-line to isolate isotopic reservoirs in the absence of a planet. }

\titlerunning{Pressure maxima inward the snow-line}

\author{S.~Charnoz \inst{\ref{inst1}} \and G.~Avice \inst{\ref{inst1}}\and
R.~Hyodo\inst{\ref{inst2}} \and F.~C.~Pignatale\inst{\ref{inst1}} \and M.~Chaussidon \inst{\ref{inst1}}}

\institute{Université de Paris, Institut de physique du globe de Paris, CNRS, F-75005 Paris, France \label{inst1} \and  ISAS/JAXA, Tokyo, Japan \label{inst2}
}

\date{\today}

 \abstract{ Pressure maxima are regions in protoplanetary disks where pebbles can be trapped because of the local absence of pressure gradient. These regions could be ideal places to form planetesimals or to isolate isotopic reservoirs. Observations of protoplanetary disks show that dusty rings structures are common, and pressure maxima are sometime invoked as a possible explanation. In our Solar System, pressure bumps have been suggested as a  possible mechanism for separating reservoirs with different nucleosynthetic compositions, identified among chondrites and iron meteorites. In this letter we detail a mechanism by which pressure maxima form just inward the snow-line in stratified disks (with a dead-zone and an active layer). This mechanism does not need the presence of a planet. }
{We investigate the conditions for pressure maxima formation using a vertically averaged $\alpha$ viscosity model and release of water vapor at the snow-line.}
{We consider 1D-$\alpha$ disk model. Using a combination of analytical and numerical investigation we explore the range of conditions for a pressure maximum to form inside the dead-zone and just inward the snow-line.}
{When the vertically averaged $\alpha$ is a decreasing function of surface density then the release of water vapor at the snow-line lowers the sound velocity, and in turn, a pressure bump appears. This requires a constant inflow of icy pebbles with pebbles influx to gas influx $>0.6$ for a  power law disk with $1\%$ ice/gas ratio, and $>1.8$ for a disk with ice/gas ratio $\sim 0.3\%$.If these conditions are met, then a Pressure-maximum appears just inward the snow-line due to a process coupling the dead and active layers at the evaporation front. The pressure bump survives as long as the icy pebble flux is high enough.The formation of the pressure bump is triggered by the  drop of sound velocity inward the snow-line, due to the release of water vapor. }
{This mechanism is promising for isolating early reservoirs carrying different isotopic signatures in the Solar System and for promoting dry planetesimal formation inward the snow-line, provided the vertically averaged description of a dead-zone is valid.}

\keywords{protoplanetary disks, planetesimal formation, $\alpha$ disk models.}

\maketitle
\section{introduction}
Formation of pressure maxima in protoplanetary disks is an active topic of research because they are seen as ideal places where pebbles could accumulate efficiently, and subsequently form planetesimals through, for example, the so called "Streaming instability" process \citep{Johansen_2007, Pinilla_2012, drazkowska_2014,drazkowska_2017, charnoz_2019}. Pressure bumps could also act as dynamical barriers where particles with Stokes-number (St) that deviates from 0 could have their radial drift slowed-down, stopped, or even reversed, because the disk becomes super-keplerian for a positive pressure gradient. Dust and pebbles experience gas-drag. Their drift velocity (relative to the gas) is  \citep{Weidenschilling_1977}:

\begin{equation}
V_{drift}=\frac{St}{2\Omega\rho_g (1+St^2)} \frac{dP}{dr} 
\label{eq_Vn}
\end{equation}

where $\Omega$ is the local keplerian frequency, $\rho_g$ the gas density and $P$ the gas pressure. Since the direction of migration (relative to the gas) is dictated by $\partial P / \partial r$, a pressure bump is defined as a pressure maximum, so that on either sides, drifting pebbles moves toward the local pressure maximum. Conversely, for a pressure minimum, drifting pebbles will go away from the minimum.

For these reasons, a pressure maximum is also invoked as a possible dynamical barrier that could be at the origin  of a major isotopic heterogeneity observed in Solar System meteorites: Meteorites can be divided in two broad groups on the basis of their non mass-dependent variations of stable isotopes (\cite{Trinquier_2008, Kruijer_2017, Kruijer_2020, Kleine_2020}). These are: the Carbonaceous Chondrites (CC) group and the Non-Carbonaceous chondrites (NC) group. Iron meteorites, as well, from their isotopic anomalies, can also be partitioned into the CC and NC groups. Accretion timescales modeled from Hf/W ages of metal silicate differentiation in parent bodies of iron meteorites belonging to the NC and CC groups show that NC bodies accreted earlier than CC bodies with however an overlap at around $\simeq 1.4 Myr$ after CAIs. Accretion ages of parent bodies of chondrites and iron meteorites show that the NC and CC reservoirs were isolated very early in the disk (<0.4Myr after CAIs for the NC group) and during nearly 2 Myrs \citep{Kruijer_2017, Kleine_2020}.

Jupiter is often invoked \citep{Kruijer_2017, Kleine_2020} as responsible for separating the two reservoirs, however it was recently argued that Jupiter may form too late for the isolation process to be efficient enough \citep{Brasser_2020}. In the same study, it is proposed that the isolation may result from the presence of a pressure maximum, appearing in a few 100Kyrs after CAIs, however, without detailing how this pressure bump could have formed. In addition, the very early model age for the accretion of the parent bodies of NC iron meteorites (<0.4 Myrs after CAIs, \cite{Kruijer_2017}) shows that the separated NC and CC reservoirs existed in the disk probably before the formation of Jupiter.

Pressure maxima could also be interesting mechanisms explaining other observed disk features such as dusty ringed structures in transition disks (see e.g. \citet{Pinilla_2012, Van_der_Marel_2018}) or as means to concentrate dust to form water-poor planetesimals if their formation occurred $\textit{inward}$ the snow-line \citep{Ida_Guillot_2016, Ida2021, charnoz_2019, Hyodo_2019, Hyodo2021a}.

How pressure bump could form in the absence of a planet is still unclear and the present study is an attempt to answer this question. Charnoz et al. (2019) reports the formation of a long-lasting pressure bump just \textit{inward} the snow-line and the subsequent runaway accumulation of dust at the bump, when considering a stratified accretion disk with a dead zone. However, the mechanism triggering this bump was not thoroughly investigated, but was clearly associated to the release of water vapor inward the snow-line. Here we try to understand how a bump could form inward a snow-line, using the popular $\alpha$ disk description (which is not devoid of problems also, see e.g. \citet{Turner_2014} for a critical review). 

In section 2 we introduce a simple parameterization of $\alpha$ to capture the effect of a dead zone in a protoplanetary disk. We detail with simple analytical arguments how the release of water vapor will trigger the formation of a pressure bum. In section 3 we demonstrate the existence of process with numerical simulations and investigate the space of free parameters. In section 4 we investigate at what epoch a pressure bump could form during the disk evolution. Our results are discussed in section 5 with a special emphasis on the separation of isotopic reservoirs as observed in our Solar System.

\section{1D Analytical study of a disk with a $\alpha$ decreasing with surface density}
\subsection{Vertically averaged $\alpha$}
 Protoplanetary disks are thought to have a vertically stratified structure due to non-ideal MHD effects. The disk's midplane may have a low turbulence, and low accretion rate, due to ohmic diffusion that prevents the onset of magneto-rotational instability \citep{Turner_2014}. This results in a quiet 'dead' midplane with equivalent $\alpha$ (called $\alpha_d$) in the range $10^{-5}$ to $10^{-3}$ depending on  local  hydrodynamic turbulence \citep{Bai_Stone_2013, Turner_2014, Gressel_2015, Kadam_2019}. It is topped by an active layer with high accretion rate, but with a low column-density ($\Sigma_a$), in the range 100 to 1000 $kg/m^2$ \citep{Turner_2014} and that may have a low level of turbulence (\cite{Bethune_2017}) despite of a high accretion rate. In the active layer the effective value of $\alpha$ is designated by $\alpha_a$. The very upper layer may be occupied by disk-winds, a region in which ambipolar diffusion is active, that torques the two previous layers and where low-density winds breeze outwards. The transition between the dead and active layers (in terms of column density) may be very sharp \citep{Turner_Sano_2008, Bai_Stone_2013, Turner_2014}. In 1D models it  is useful to introduce the $\textit{vertically averaged $\alpha$}$ :

 \begin{equation}
\left<\alpha\right>= \frac{1}{\Sigma}\int_{-\infty}^{+\infty} \rho(z) \alpha(z) dz
\label{equation_v_drift}
\end{equation}   

where $\rho(z)$ and $\alpha(z)$ are the values of the gas density and $\alpha$ at altitude z respectively. As the transition of the active to the dead layer is very sharp, it is possible to use a parameterization of $\left<\alpha\right>$ assuming that the disk is made of  two layers: an active layer with constant column density $\Sigma_a$ (100 to 1000 $kg/m^2$, largely independent of the distance to the star, \cite{Turner_Sano_2008, Turner_2014}) and constant $\alpha=\alpha_a$, and a dead midplane layer with surface density $\Sigma_d=\Sigma-\Sigma_a$ and constant $\alpha=\alpha_d$ (see Appendix). Thus we get \citep{Terquem_2008, Zhu_2010, Bai_Stone_2013, charnoz_2019, Kadam_2019}:

\begin{equation}
\left\{
\begin{array}{ll}
\text{if $\Sigma \geq \Sigma_a$ :} & \quad \left<\alpha\right> (\Sigma)= \frac{\Sigma_a \alpha_a+(\Sigma-\Sigma_a)\alpha_d}{\Sigma}  \\
\text{if $\Sigma < \Sigma_a$ :}  & \quad \left<\alpha\right> (\Sigma)= \alpha_a
\end{array}
\right.
\label{eq_alpha_dz}
\end{equation}

Following \cite{Kadam_2019} we use $\alpha_a=10^{-2}$ and $\alpha_d=10^{-5}$. In the following, to make the discussion clear, we use a simpler parameterization of $\left<\alpha\right>$ so that $\left<\alpha\right> \propto \Sigma^{-p}$. So it is useful to introduce $p$ the  exponent of the power-law locally approximating $\left<\alpha\right>(\Sigma)$, it is :
\begin{equation}
p=\frac{-\Sigma}{\left<\alpha\right>}\frac{\partial \left<\alpha\right>}{\partial \Sigma}
\label{power_law_exponent}
\end{equation}

\begin{figure}
	\includegraphics[scale=0.30]{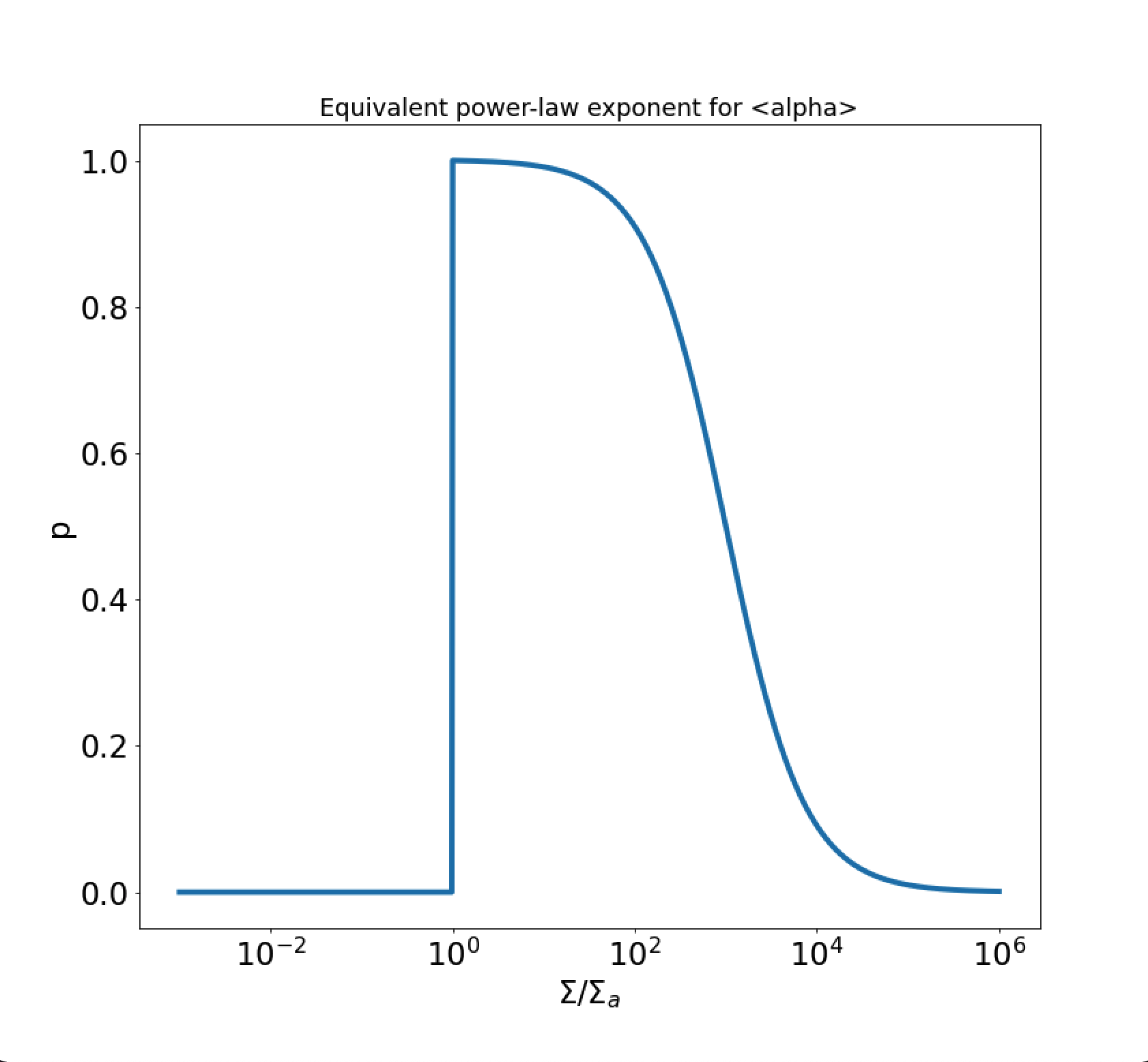}
 \caption{$p$, as a function of $\Sigma$  for $\left<\alpha\right>$ as defined in Equation \ref{eq_alpha_dz}. $p$ is defined  in Equation \ref{power_law_exponent} }
 \label{Figure_equivalent_p}
\end{figure}
Inserting Equation \ref{eq_alpha_dz} into \ref{power_law_exponent} we find that p ranges from 0 to $(\alpha_a - \alpha_d) / \alpha_a$. Since $\alpha_a >> \alpha_d$, p can get very close to 1.
$p$ is plotted in  Figure \ref{Figure_equivalent_p}. For $\Sigma < \Sigma_a$, p=0 because $\left<\alpha\right>$ is a constant equal to $\alpha_a$. For $\Sigma > \Sigma_a$ and  $\Sigma < \Sigma_a \alpha_a/\alpha_d $ p decreases from $\sim 1$ to 0. For  $\Sigma > \Sigma_a  \alpha_a/\alpha_d $ p is almost equal to 0. \\

\label{section_alpha}
\subsection{Small amplitude perturbation}
For illustrative purpose we start by considering the case of a small sound velocity perturbation, to emphasize the strong effect it has on the disk's pressure profile.
The disk is described through the standard $\alpha$ disk formalism, with gas surface density $\Sigma$ evolution obeying to :
\begin{equation}
\frac{\partial \Sigma}{\partial t} = \frac{3}{r}\frac{\partial}{\partial r}\left( r^{1/2} \frac{\partial}{\partial r}(\nu \Sigma r^{1/2})\right) 
\label{Eq_Sigma_evol}
\end{equation}   
where $\nu$ and $r$ are the local viscosity and distance to the star.
Steady states are those solutions for which $\partial \Sigma / \partial t=0$ implying $\nu \Sigma=cst$. Noting that the mass flux $\dot M = 3 \pi \nu \Sigma$, we get at steady state  $\Sigma = \frac{\dot M }{3 \pi \nu}$, which is well known. The effective viscosity is $\nu=\left<\alpha\right>C_s^2/\Omega$ where $C_s$ is the local sound velocity. We assume that $\left<\alpha\right>$ can be written:
\begin{equation}
 \left<\alpha\right>=A\Sigma^{-p}
 \label{eq_alpha-P}
 \end{equation}  

where A is a constant and p is a positive real number $\leq 1$. Note that for every-value of $\Sigma$ there is a different value of p.

We replace Equation \ref{eq_alpha-P} into the expression of the steady state surface density: $\dot M = 3 \pi \nu \Sigma$. After solving for $\Sigma$ we get the surface density in the midplane at steady state $\Sigma_{ss}$ inside a dead-zone and for a fixed value of p and $\Sigma$:

\begin{equation}
 \Sigma_{ss}=\left( \frac{\dot M  \Omega }{3 \pi A C_s^2} \right)^{\frac{1}{1-p}}
 \label{sigma_ss}
 \end{equation}  

It is obvious from Equation \ref{sigma_ss} that  $\Sigma_{ss}$ has no steady-state solution for p=1, and for p<1 we see that a minimum of sound velocity induces a maximum of surface density. Note that in the above equation p is fixed (given by Equation \ref{power_law_exponent}).
For locally isothermal 
pressure  $P=\Sigma \Omega C_s/(2\pi)^{1/2}$, we get for the local pressure at steady state $P_{ss}$:

\begin{equation}
 P_{ss}=\frac{\Omega^{\frac{2-p}{1-p}}}{(2 \pi)^{1/2}}\left( \frac{\dot M  }{3 \pi A} \right)^{\frac{1}{1-p}}C_s^{\frac{-(1+p)}{1-p}}
 \label{p_ss}
 \end{equation} 
We now quantify the effect of a sound velocity perturbation ($\delta C_s$) on the pressure profile at steady state. Such a sound velocity perturbation may result from a local variation of gas-mean molecular weight  because of the release of water vapor at the snow-line. By doing a first order expansion of $P_{ss}$ as a function of $C_s$ we get $\delta P_{ss}$ the pressure perturbation at steady state, as a function of the sound velocity perturbation $\delta C_s$:
\begin{equation}
    \delta P_{ss}=-\frac{1+p}{1-p}P_{ss}\frac{\delta C_s}{C_s}
\end{equation}

 Again we see that the amplitude of the pressure perturbation diverges when $p \rightarrow 1$, and that $\delta P_{ss}$ has a sign opposite to the sound-velocity perturbation. We now determine how strong must be $\delta C_s$ to induce a pressure bump by taking the derivative of equation \ref{p_ss} and setting that for a pressure bump $\partial P_{ss} /\partial r \geq 0$. We write $\Omega=Br^{-3/2}$ (with $B$ a constant) and we obtain the condition for a pressure bump to appear by deriving equation \ref{p_ss} with respect to $r$:

\begin{equation}
-\frac{3 a}{2 r}-\frac{b}{C_s}\frac{\partial C_s}{\partial r} \geq 0
\label{Cs_cond}
\end{equation} 
with $a= (2-p)/(1-p)$ and $b=(1+p)/(1-p)$ (two positive constants). Thus a pressure bump appears if the local derivative of $C_s$ verifies $ \partial C_s/\partial r \leq
-3a/2br$, which is a negative sound velocity perturbation. 

Assuming that the disk is radiatively heated by the star, the unperturbed sound velocity profiles behaves like  $C_r=D r^{-1/4}$ , with D standing for a positive constant. We compute the amplitude of the sound velocity perturbation $\delta C$, above the  unperturbed radiative profile, $C_r$, to trigger a pressure bump. We simply write $C_s=C_r+\delta C$ and introduce it in equation \ref{Cs_cond}. We get the amplitude of the sound velocity perturbation to induce a pressure bump :

  \begin{equation}
\frac{\partial \delta C}{\partial r} \leq -\frac{C_r}{r} \frac{(11-7p)}{4(1+p)}
\label{criter_bump_ss}
 \end{equation} 
 for \textit{p} close-to, but smaller than, 1, we get $-\frac{\partial \delta C}{\partial r} \geq \frac{1}{2}\frac{C_r}{r}$. So the possibility to form a pressure bump depends both on the amplitude of the sound perturbation ($\delta C$) and on its width ($ \delta r$) by approximating $\frac{\partial \delta C}{\partial r} \simeq   \delta C/\delta r$. 
 In the present paper we are interested in sound-velocity perturbations induced by the release of water vapor just inward the snow-line that modify the mean-molecular weight of the gas at constant temperature. As the isothermal sound velocity is $C=(RT/\mu)^{1/2}$ we get $\delta C=-1/2C \delta \mu/\mu$. So Equation \ref{criter_bump_ss} is simply rewritten :
 
 \begin{equation}
\frac{\delta \mu}{\mu} \geq \frac{\delta r}{r} \frac{(11-7p)}{2(1+p)}
\label{criter_bump_ss_mu}
 \end{equation} 
 
 $H_2O$ has a molecular weight $\mu_{H2O}=18g/mol$ and the average gas of solar composition has $\mu_g=2.3g/mol$. The mean molecular weight is $1/\mu=f/\mu_{H2O}+(1-f)/\mu_g$ \citep{Schoonenberg_2017}. Noting $f$ the water mass fraction, we get $\delta \mu/\mu=\delta f \mu(1/\mu_g-1/\mu_{H2O})  \simeq 0.87 \delta f $. So the previous equation is rewritten in term of water mass fraction :
 
 \begin{equation}
\delta f\geq \frac{\delta r}{r} \frac{2(11-7p)}{3.5(1+p)}
\label{criter_bump_ss_f}
 \end{equation}
 
\begin{figure}
	\includegraphics[scale=0.40]{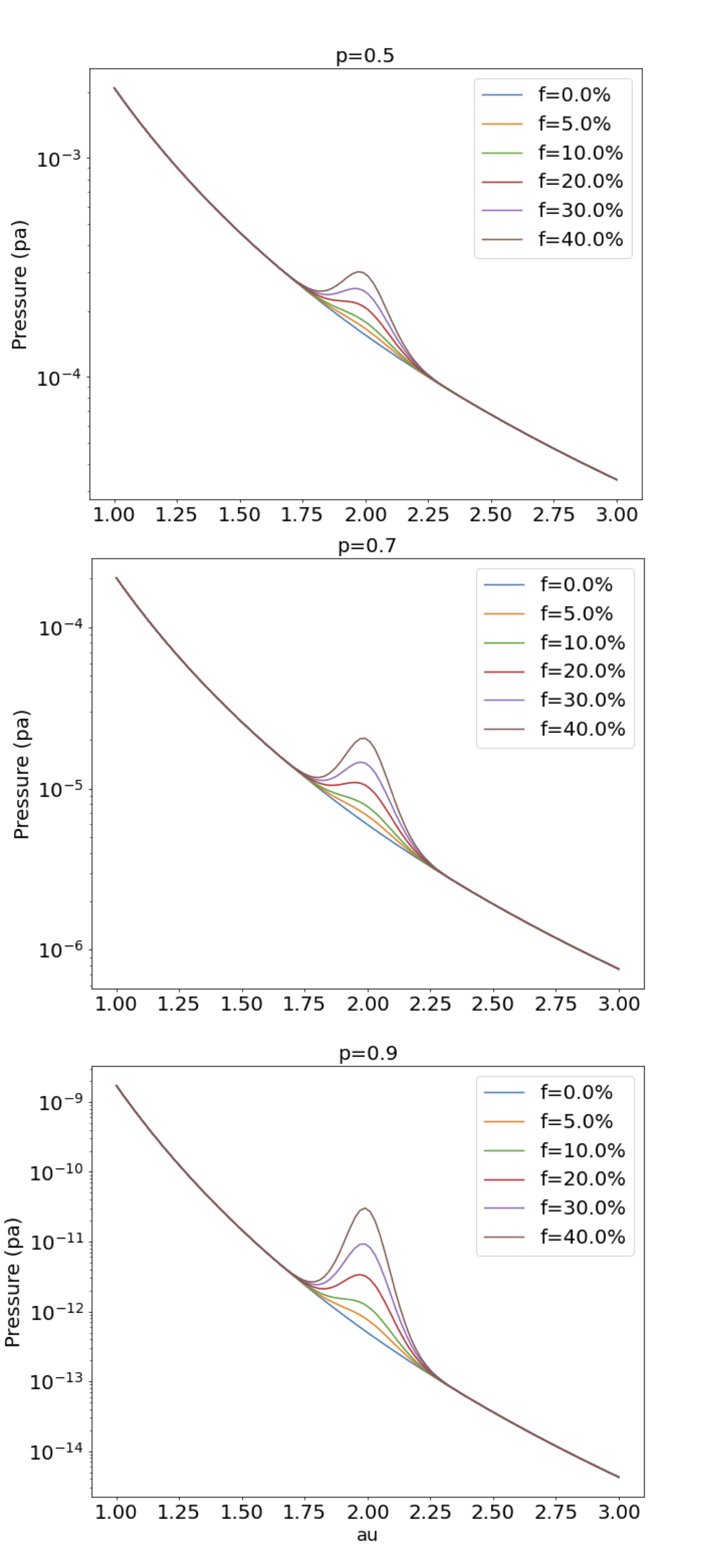}
 \caption{Pressure Vs. distance in disks at steady states with different mass water-vapor fractions ($f$) enhancement at the snow-line  (located at 3 au here). Here $\dot M=10^{-9} M_{\odot}/year$ and p=0.1 (top), p=0.5 (middle), p=0.9 (bottom). The width of the water-vapor rich region is arbitrarily set to 0.1 au (numerically and analytically derived in \citet{Hyodo2021a}). For a same mass fraction of water vapor, the pressure bump's amplitude increases for larger p. }
  \label{Figure_bump_SS}
\end{figure}

 Here $\delta r$ is the radial spatial scale of variation of water vapor content, typically the physical width of the snow-line controlled by the saturating vapor. $\delta r$ is derived in \cite{Hyodo2021a} and $\delta r$ is about $\sim 0.2$ AU (their Equation 54) with the model settings used in \cite{Schoonenberg_2017}. $\delta f $ is the amplitude of variation of the water vapor mass fraction. Before the inflow of water ice coming from beyond the snow-line $f\simeq 0.01$ that is the average water mass fraction in a disk of solar composition. Equation \ref{criter_bump_ss_f} provides a condition for the apparition of a pressure bump in the presence of increasing water-vapor at the snow-lines. It states that a pressure bump will appear if $\delta f \geq  0.28$ for p=0.5 or $\delta f \geq  0.14$ for p=1.  To test this criterion, we have done some simple tests  with an $\alpha$ disk model solving Equation \ref{Eq_Sigma_evol} for gas and computing the viscosity using Equation \ref{eq_alpha-P}, so that p can be fixed. The initial water mass fraction, $f$, and $\delta r$, are set by hand using a gaussian profile with standard deviation 0.1 au \citep{Hyodo2021a}. When Equation \ref{criter_bump_ss_f} is satisfied, we do see the formation of a pressure bump with this "toy" model. A pressure maximum appears for $\delta f\simeq 0.2$ in Figure \ref{Figure_bump_SS}. Of course the surface density profiles changes also slightly according to Equation \ref{sigma_ss}. So in for p close to 1, we will keep the criterion for pressure bump formation as $\delta f > 0.2$
 
 \subsection{Large amplitude perturbation}

 We have considered above for illustrative purpose, only a small perturbation of the sound velocity, and thus a small amount of water vapor. But it is found that the sound velocity perturbation necessary to induce a bump is, in fact, not so small (since $-\frac{\partial \delta C}{\partial r} \geq \frac{1}{2}\frac{C_r}{r}$). So, to continue our demonstration we abandon our first order development and go back to the original equations. We first compute the steady-state surface density by inserting Equation \ref{eq_alpha_dz} into the steady state flux $\dot M = 3 \pi \nu \Sigma$. As it is common to parameterize the disk's evolution using the accretion rate, $\dot M$, we have computed for different $\dot M$ the steady state disk surface density assuming the dead-zone prescription for $\left<\alpha\right>$ (Equation \ref{eq_alpha_dz}) combined with the steady-state relation  $\dot M=3\pi \nu \Sigma$. After a little algebra one finds the steady state surface density $\Sigma_{ss}$:

\begin{equation}
\left\{
\begin{array}{ll}
\text{if $\dot M \geq \dot M_{DZ}$ :} & \Sigma_{ss}= \frac{\dot M \Omega}{\alpha_d 3 \pi C_s^2}+\Sigma_a\left( 1-\frac{\alpha_a}{\alpha_d}\right)  \\
\text{if $\dot M < \dot M_{DZ}$ :}  & \Sigma_{ss}= \frac{\dot M \Omega}{\alpha_d 3 \pi C_s^2} 
\end{array}
\right.
\label{eq_sigma_steady_dz}
\end{equation}

where $\dot M_{DZ}$ is the critical accretion rate beyond which a region is embedded in dead-zone : $\dot M_{DZ}=\Sigma_a \alpha_a 3 \pi C_s^2/\Omega$.

 Assuming $\dot M > \dot M_{DZ}$, the pressure in the Dead Zone at steady state is then :
 
 \begin{equation}
     P=\frac{1}{(2 \pi)^{1/2}}\left( \frac{\dot M \Omega^2}{\alpha_d 3 \pi C_s}+\Omega C_s \Sigma_a (1-\alpha_a/\alpha_d) \right)
 \end{equation}
 
 Then, we simply determine the condition for which $\partial P/ \partial r \geq 0$ in the DZ. We get :
 \begin{equation}
     \frac{\partial C_s}{\partial r} \leq \frac{-3 C_s}{ r} \left(\frac{A_1-A_2/2}{A_1+A_2}  \right)
     \label{criter_bump_ss_Cs}
 \end{equation}
 with coefficients:
 \begin{equation}
 A_1=\frac{\dot M \Omega^2}{\alpha_d 3 \pi C_s}
 \end {equation}
 \begin{equation}
 A_2=\Omega C_s \Sigma_a (\alpha_a/\alpha_d-1)
 \end {equation}
 We now try to reformulate the above criterion in terms of water mass fraction enhancement at the snow-line. Since $C_s=(RT/\mu)^{1/2}$, then
  $ dC_s=1/2 C_s dT/T-1/2 C_s d\mu/\mu$. In the region at the snow-line where water vapor is released we assume that $d\mu/\mu >> dT/T$ so  $ dC_s \simeq-1/2 d\mu/\mu$ \citep{Hyodo2021a}. Since $\delta \mu/\mu \simeq 0.87 \delta f $, the bump formation criterion (Equation \ref{criter_bump_ss_Cs} is rewritten in terms of water vapor mass fraction ($\delta f$):
 
  \begin{equation}
     \delta f > 7 \frac{\delta_r}{ r} \left(\frac{A_1-A_2/2}{A_1+A_2}  \right)
     \label{criter_bump_ss_true}
 \end{equation}
 
\begin{figure}
\includegraphics[scale=0.4]{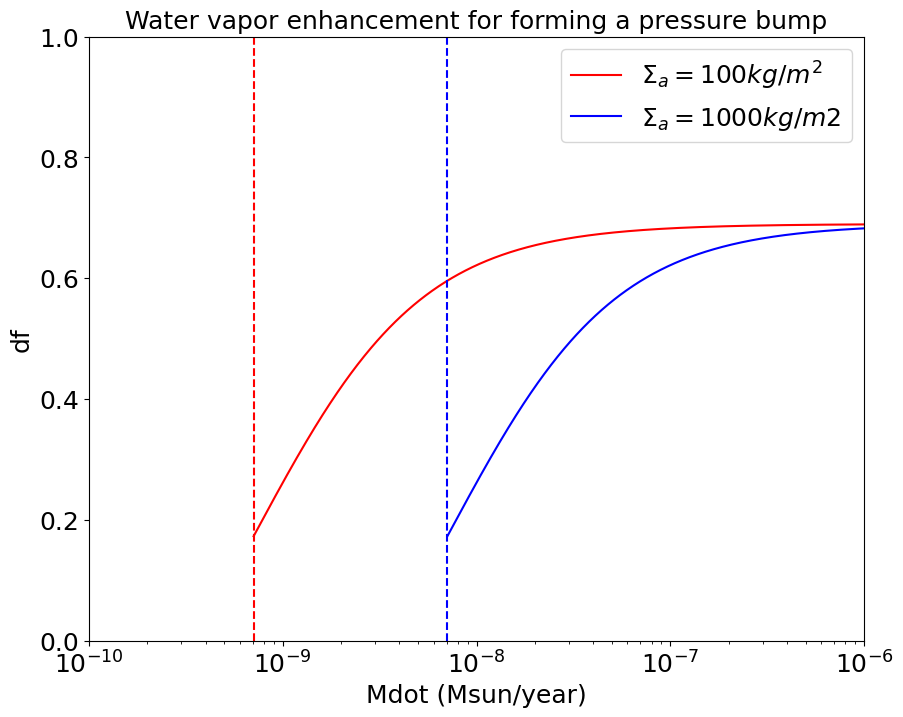}
 \caption{Water vapor fraction at the snow-line necessary to induce a pressure bump, as a function of the accretion rate ($\dot M$), at 2 au. Red solid line $\delta f$ for :$\Sigma_a=100 kg/m^2$, blue solid line  $\Sigma_a=1000 kg/m^2$. Dashed red and blue line display $\dot M_DZ$, the accretion rate for which the 2au region would be in a dead-zone. }
  \label{Figure_df_VS_Mdot}
\end{figure}

 The above equation gives a much better criterion for forming a pressure bump even in the case of strong sound velocity perturbation. The water vapor enhancement necessary for forming a pressure bump, $\delta f$ as a function of $\dot M$ is plotted in figure \ref{Figure_df_VS_Mdot} for $\delta r/r \simeq 0.1 $. At high surface densities and high accretion rate $\delta f \simeq 0.7$ (equivalent to $A_2 << A_1$ in Equation \ref{criter_bump_ss_true}). A low accretion rate, $\delta f \simeq 0.7$ in agreement with estimates of the previous section. 
 So it is interesting to see that, as the accretion rate decreases, it is increasingly easy for the disk to develop a pressure bump (as $\delta f$ drops from 0.7 to 0.2). Once the disk has lost most of its mass, and once the snow-line is no longer inside the dead-zone it is no longer possible to develop a pressure bump.
 
 For the high accretion rates, relevant to young disks, we will keep, for simplicity purpose, the criterion :
  \begin{equation}
     \delta f > 7 \frac{\delta_r}{ r}
  \end{equation}

 \subsection{Criterion in terms of icy pebbles mass flux}
 
The criterion derived in Equation \ref{criter_bump_ss_mu} does not allow to predict the range of pebble and gas flux for which a pressure bump may form inside a dead-zone. Thus, we propose the following order of magnitude estimate : the vaporization region (the region around the snow-line where icy pebbles evaporate) is located at distance $r$ and with width $\delta r$. The water mass contained in this region is $M_w$. It is fed by the incoming flux of icy pebbles $F_i=2 \pi r v_i \Sigma_i$ with ice surface density $\Sigma_i$ and icy pebbles velocity $v_i$. Water vapor leaves this region with the same velocity as the gas $v_g$ so the water vapor flux leaving the region is $F_v=2 \pi r v_g \Sigma_w $. After a transient phase, the surface density of water vapor will reach a steady state, with surface density at steady-state $\Sigma_{w,ss}=\Sigma_i v_i/v_g$. For a pressure bump to appear we must have $\Sigma_{w,ss}/\Sigma > \delta f$. So we get another (equivalent) condition for a pressure bump to form, as a function of gas velocity and ice surface density :

\begin{equation}
    \frac{v_i}{v_g}>7 \frac{\Sigma}{\Sigma_i} \frac{\delta r}{r} 
    \label{eq_criterion_vd_vg}
\end{equation}
for $\delta r \simeq 0.2$ au \citep{Hyodo2021a} and $r \simeq 2.0 $ au and for $\Sigma_i/\Sigma  \simeq 0.01$ (as typical values) we get $v_i/v_g > 70$ for a pressure bump to appear. Equivalently Equation \ref{eq_criterion_vd_vg} can be formulated in terms of pebble/gas flux:

\begin{equation}
    \frac{F_i}{F_g}>7 \frac{\delta r}{r}
    \label{eq_criterion_Fi_Fg}
\end{equation}

and with the above values we get $F_i/F_g > 0.7$.

In summary, the conditions for  pressure bump to appear at the snow-line are :
\begin{itemize}
    \item The snow-line must be embedded  within the dead-zone (and \textit{not} necessarily close to its edge).
    \item The ice flux necessary for form a pressure bump depends on the surface density at the snow-line. The higher the surface density (for young disks with high accretion rates), the higher the necessary flux. Conversely, as surface density decreases, as as the accretion rate decreases also, the lower the necessary ice flux to form a pressure bump.
    
    \item For high surface density disk, corresponding to high accretion rate ($>10^-7 M_{\odot}/year$), assuming the $\Sigma_i/\Sigma \simeq 0.01$, a  pressure bump will form at the snow-line if  $v_i/v_g$  > 70, or equivalently, if $F_i/F_g > 0.7$ for an ice to gas ratio of $1\%$.  This corresponds to $p$ close to 0. So a high pebble flux is needed in dense disks to form a pressure bump.
    \item For low surface density disk ( and accretion rate in the range$ \simeq 10^-9 M_{\odot}/year$), assuming the $\Sigma_i/\Sigma \simeq 0.01$, a  pressure bump will form at the snow-line if  $v_i/v_g$  > 20, or equivalently, if $F_i/F_g > 0.2$ for an ice to gas ratio of $1\%$. This corresponds to case with $p$ close to 1, so it is easier in low surface density dead-zones to form a pressure bump.

\end{itemize}

Note that the above calculations assume that the surface density of the disk has reached a steady state (i.e. constant accretion rate everywhere). However, simulations of disks with including Dead Zone \citep{Zhu_2010,Hasegawa_Takeuchi_2015, charnoz_2019} show that in general, a steady state is never reached, and that the disk suffers episodic phases of high and low accretion rates. So in the next section, we study numerically the formation of a pressure bump, for a disk which is not at steady state.

\section{Comparison with time evolving simulations}

\begin{figure}
	\includegraphics[scale=0.30]{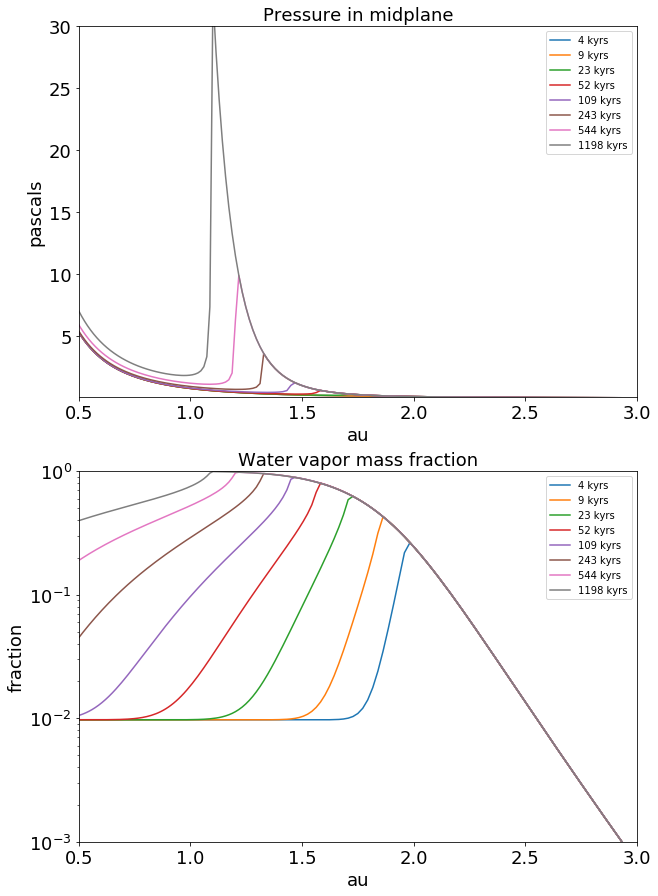}
\caption{Evolution of a disk with constant inflow of gas and pebbles. Here $\Sigma_a=100 kg/m^2$ and pebbles Stokes number=0.1. (top): Pressures in the disk midplane vs. distance at different times (bottom) water vapor mass fraction vs. distance for different times. See table \ref{table_ST100}. }
  \label{fig_disk_evol_sigma1000_st01}
\end{figure}

We have run 12 simulations of disks evolution, in order to check the validity of the criteria established in the previous section  that assume accretion at steady state so that the surface density follows Equation \ref{eq_sigma_steady_dz}. Here, we assume a power-law surface density disk, so that is not at steady state, but correspond to a more realistic case compared to the previous section. We implement a simple time evolving alpha disk model very similar to the one described in \cite{Schoonenberg_2017} and  \cite{Hyodo_2019} to treat gas, pebbles and water vapor through diffusion and advection. Note that  the influx of icy pebbles is kept constant as well as the Stokes number of pebbles during their drift. The temperature profile is also constant and decreases like $r^{-1/2}$. Evaporation of water vapor is treated as in \cite{Schoonenberg_2017} and results in a change of the local sound velocity that change the effective viscosity $\nu$ through the relation $\nu=\left<\alpha\right>C_s^2/\Omega$. The initial surface density of gas follows  $\Sigma(r)=10^4(r/1 au)^{-1} kg/m^2 $ and the temperature profile is $T(r)=150K(r/2 au)^{-1/2}$.  The only difference with \cite{Schoonenberg_2017} is the computation of $\alpha$ where we use the dead-zone prescription (Equation \ref{eq_alpha_dz}) with $\alpha_d=10^{-5}$ and $\alpha_a=10^{-2}$. From one simulation to another the Stokes number is varied from 0.01 to 0.1. The accretion rate is close to $10^{-7} \dot M_{odot}/year$, so that the the theoretical criterion for forming a pressure bump is that the ice flux must be > 0.6-0.7 times the gas flux. In the first set of simulations (Table \ref{table_ST100}) $\Sigma_a$ is set to $100 kg/m^2$ so that the outer edge of the dead-zone is located at 100 $au$ whereas the snow-line is initially located around 2 $au$. We present in Figure  \ref{fig_disk_evol_sigma1000_st01} the case with St=0.1  In this configuration the ratio $v_i/v_g=822.1$. The formation of a pressure bumps appears at about 6 Kyrs (Table \ref{table_ST100}). We see that the pressure bump never disappears and shifts inward by about 1 au as water vapor accumulates and increases the  sublimation temperature. Like in \cite{Schoonenberg_2017}
we do see that the water vapor mass fraction (f) evolves toward an asymptotic steady state where the water vapor mass fraction tends to decrease monotonically with distance. However, even after $\>1$ Myrs evolution this steady state is still not reached here because of low effective viscosity of the gas in the disk midplane. This represents a substantial fraction of the disk lifetime. Material tends to pile up at the snow-line with increasing efficiency as the viscosity drops with incoming water vapor, resulting in a sharper pressure bump with time.

In all simulations we could perform for the case $\Sigma_a=100 kg/m^2$ we observe the formation of a pressure bump as all our runs satisfy the criterion of Equation \ref{eq_criterion_vd_vg}, with $v_i/v_g> 84$ for Stokes number as small as 0.01. Because of expansive simulation time we could not investigate smaller values of $v_i/v_g$ for this kind of run. In a second set of runs, we set $\Sigma_a=1000kg/m^2$ and we find formation of a pressure bump down to pebbles Stokes number=0.075 (Table\ref{table_ST1000}) and $v_i/v_g=66.4$ and $F_i/F_g = 0.6$. For lower values of the Stokes number, and for $v_i/v_g<66$ we do not find any pressure bump formation up to 8.5 Myrs evolution of the disk. So it seems that our above simple reasoning gives a very reasonable estimate of the threshold. So in the rest of our discussion we will adopt $v_i/v_g > 60 $ as the criterion for pressure bump formation for simplicity (and equivalently $F_i/F_g > 0.6$, assuming a water to gas ratio of $1\%$). We emphasize that in these simulations gas diffusion, included in the calculation, is indeed acting against the piling up of gas to smooth radially the surface density peak. However, since $\alpha$ is a decreasing function of surface density ( Equation \ref{eq_alpha_dz}) diffusion is less and less efficient as the surface density increases. As a result surface density peaks are enhanced.


\begin{table*}
\begin{tabular} {|l|l|l|l|l|l|l|}
  \hline
  \textbf{$\Sigma_a=100 kg/m^2$}   & \textbf{St=0.01} & \textbf{St=0.02} & \textbf{St=0.05} & \textbf{St=0.075} & \textbf{St=0.1}  & \textbf{St=0.2} \\
  \hline
  $v_i$  &  -0.80 m/s & -1.60 m/s & -3.96 m/s  & -5.92 m/s &  -7.85 m/s  & -15.3 m/s  \\
  $v_g$  &  -0.0095 m/s & -0.0095 m/s & -0.0095 m/s & -0.0095 m/s &  -0.0095 m/s & -0.0095 m/s \\
  $v_i/v_g$ &  84.6 & 167.4    & 415. & 619.7 & 822.1 & 1595.1  \\
  
  $F_i/F_g$ &  0.8 & 1.6    & 4.1 & 6.2 & 8.2 & 15.9  \\
  
  \textbf{Pressure Bump} ? & \textbf{YES} & \textbf{YES} & \textbf{YES} & \textbf{YES} & \textbf{YES} & \textbf{YES} \\
  
  Time PB formation & 1.45 Myrs  & 300 Kyrs    & 25 Kyrs    &    21 Kyrs & 6 Kyrs &  2.7 Kyrs \\

  \hline

\end{tabular} 
\caption{\label{table_ST100} Summary of results obtained for 6 simulations with $\Sigma_a=100 kg/m^2$. "Time of PB formation" is the time at which the pressure maximum appears. All these simulations lead to the formation of a pressure bump, but at different times.}
\end{table*}

\medskip

\begin{table*}
\begin{tabular} {|l|l|l|l|l|l|l|}
  \hline
  \textbf{$\Sigma_a=1000 kg/m^2$}   & \textbf{St=0.01} & \textbf{St=0.02} & \textbf{St=0.05} & \textbf{St=0.075} & \textbf{St=0.1}  & \textbf{St=0.2} \\
  \hline
  $v_i$  &  -0.95 m/s & -1.74 m/s & -4.11 m/s  & -6.06 m/s &  -8.0 m/s  & -15.3 m/s  \\
  $v_g$  &  -0.091 m/s & -0.091 m/s & -0.091 m/s & -0.091 m/s &  -0.091 m/s & -0.091 m/s \\
  $v_i/v_g$ &  10.42    & 19.1 & 44.9 & 66.3 & 87.5 & 168.5 \\
  $F_i/F_g$ & 0.1 & 0.2    & 0.45 & 0.67 & 0.88 & 1.6  \\
  \textbf{Pressure Bump} ?& \textbf{NO} & \textbf{NO} & \textbf{NO} & \textbf{YES} & \textbf{YES} & \textbf{YES} \\
  Time PB formation &   &    &     & 290 Kyrs & 114 Kyrs &  21 Kyrs \\

  \hline

\end{tabular} 
\caption{\label{table_ST1000} Summary of results obtained for 6 simulations with $\Sigma_a=1000 kg/m^2$. Only simulations with peebles Stokes number $> 0.075$ lead to formation of a pressure bump. Simulations with $St<0.075$ do not lead to pressure bump formation up to 8.5 Myrscevolution."Time of PB formation" is the time at which the pressure maximum appears. }
\end{table*}

\medskip

Simulations presented in Figure \ref{fig_disk_evol_sigma1000_st01} and in Tables 1 and 2 are useful because they are simple with constant Stokes number, temperature, gas and pebble influx at the outer disk's edge. They are useful to isolate the key mechanisms. However, in real disks, the influx of icy pebbles cannot be constant, or last forever, because the ice reservoir is finite and particles grow with time. So it is interesting to see if the pressure bump may form in a more "realistic" time evolving disk including key processes : viscous spreading, non constant pebbles flux, dust and gas transport, dust sublimation and condensation, dust-growth, radiative and viscous heating. We have run a time-dependant simulation similar to \citet{drazkowska_2017} that includes the processes mentioned above (the code is described in \citet{pignatale_2018, charnoz_2019}). Note that we also use a temperature-dependant opacity table to compute the temperature in the midplane. The particle growth is computed with the model of \cite{Birnstiel_2012}. Additional details are given in \cite{charnoz_2019}.

\begin{figure*}
	\includegraphics[scale=0.32]{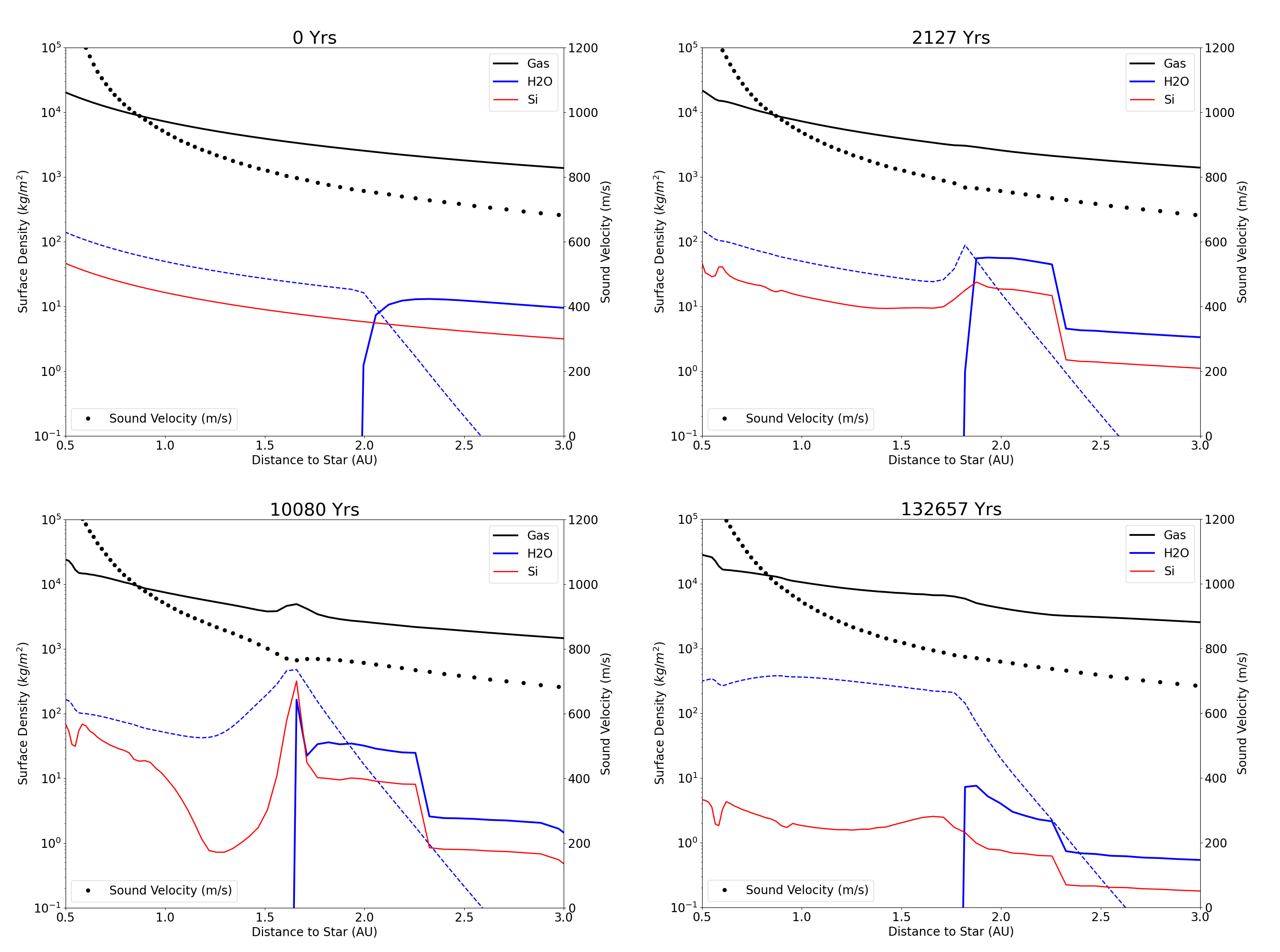}
 \caption{Simulation of a full protoplanetary disk using $\left<\alpha\right>$ given by Equation \ref{eq_alpha_dz} and running the code of \cite{charnoz_2019}. Only the region around the snow-line is displayed. Solid lines in black, blue, red represent the gas, water-ice and silicate dust surface densities (respectively), and dashed-blue line is for water vapor. The black dots display the local sound velocity in m/s (right scale).}
  \label{Figure_simu_evol}
\end{figure*}

Our initial disk has mass$=2\%M_\odot$ and $\Sigma(r)\propto r^{-1}$. We use $\left<\alpha\right>$ as defined in Equation \ref{eq_alpha_dz} with parameters $\Sigma_a=1000 kg/m^2$, $\alpha_a=10^{-2}$ and $\alpha_d=10^{-5}$. In this disk, the outer-edge of the dead-zone is about $10$ au, well beyond the snow-line (at about 2 au). With these parameters, the gas accretion rate is about $10^{-8} \dot M/year$. Results are displayed in Figure \ref{Figure_simu_evol} (it is essentially the same simulation presented in Figure 3 of \citet{charnoz_2019}). We see that at the snow-line (around 2 au) a surface density maximum forms in about 10Kyrs where a drop of sound velocity also occurs (dotted line in Figure \ref{Figure_simu_evol}). At this time the pebbles Stokes number has grown to about 0.1-0.2 just beyond the snow-line (Figure 3 of \citet{charnoz_2019}). The gas surface density maximum is accompanied  by  silicate-rich dust accumulation just \textit{inward} the snow-line that is ice-free (red line). Just \textit{outward} the snow-line, there is  also an accumulation of ice-rich dust (solid-blue line) promoted by water-vapor re-condensation and traffic jam-effect (as described in \citet{drazkowska_2014, Schoonenberg_2017, drazkowska_2017, charnoz_2019, Hyodo_2019, Hyodo2021a}). On timescales $> 100 Kyrs$ we see that the pressure bump disappears because of the emptying of the icy material reservoir beyond the snow-line. When icy pebbles flux ends the pressure-bump disappears. The water vapor surface density profile (blue dashed-line) ends close to steady-state and forms a plateau, consistently with \citet{Schoonenberg_2017} and \citet{Hyodo_2019}. Wavy structures in the silicate dust density profiles inside 1 au are due to small opacity jumps between 300K and 800K in our opacity table (see \citet{charnoz_2019}, Figure 1) that change the midplane-temperature, and thus the local viscosity. Since planetesimal formation is not considered in this simulation, nothing can save pebbles from being lost to the star ultimately when the bump disappears. If planetesimal formation was considered they should form rapidly at the location of the bump because of high-surface density and high-Stokes number. This will be investigated in a future study.

\section{Summary and Discussion}
\subsection{Summary}
We have detailed a process by which a pressure bump can form in a stratified protoplanetary disk, including a central layer dead to turbulence, and an actively accreting upper layer with surface density $\Sigma_a$. Considering that the vertically averaged $\alpha$ is a decreasing function of surface density ($\left<\alpha\right> \propto \Sigma^{-p})$, we have shown that when a local sound velocity drop appears (due to release of vapor inward the snow-line), a maximum of local pressure appears, with magnitude  increasing with $p$. A physical explanation of this process, implying an interplay between the dead and active layer, is provided in Appendix A. This effect is especially efficient when $p$ is close to 1, in other words, that the surface density at the snow-line is less than 10 times the active layer column density (which does not necessarily imply that the snow-line is located close to the outer edge of the dead-zone). Note that for a minimum mass solar nebula profile with $\Sigma(r)=1.7\times 10^4 (r/1au)^{-1.5}$ we have p>0.9 in the terrestrial planet region for $\Sigma_a> 100 kg/m^2$ (check on Figure \ref{Figure_equivalent_p}), that is already a situation very favourable for forming a pressure bump. The outer edge of the dead-zone for the same nebula is located much beyond 10 au (when $\Sigma=\Sigma_a$).

 The conditions for the pressure-bump to form are summarized below
\begin{itemize}
    \item (1) A stratified disk with a dead-zone embedding the snow-line.
    \item (2) An high influx of icy pebbles.
    \item (3) For the pressure bump to appear, a sufficient condition is that the pebble/gas velocity ratio must be $>$ 60, or in terms of ice/gas mass flux ratio the condition is  $F_i/F_g>0.6 $ for a disk with $1\%$ of water by mass.
    \item For our Solar System $1\%$ is approximately the mass fraction of all condensible material. The water mass fraction for our Solar System may be rather in the range $\sim 0.3\%$ (e.g. \cite{Bitsch_Johansen_2016}). In that case $F_i/F_g>1.8$ is needed. Such a flux is easily reached in a disk with $\Sigma_a = 100 kg/m^2$, where Stokes number $>0.03$ is needed. Conversely in a disk with   $\Sigma_a = 1000 kg/m^2$, a Stokes number $>0.3$ is needed.
    \item (4) A gas surface density at the snow-line with $ \Sigma_a < \Sigma <  \alpha_a/\alpha_d \times \Sigma_a$ (i.e. a few $10^3$ to a few $10^4 Kg/m^2$) to create the pressure bump.
    \item In the case the disk reaches a steady state accretion rate, the flux of icy particle necessary to maintain a pressure bump depend on accretion rate. It is about $F_i/F_g>0.2$ for $ 10^{-9} M_{\oplus}/year< \dot M < 10^{-8} M_{\oplus}/year$ up to   $F_i/F_g>0.6$ for $ 10^{-7} M_{\oplus}/year< \dot M$. These fluxes must be multiplied by $\sim 3$ if the ice/gas ratio is about $0.3\%$ (rather than $1\%$) like it was probably in our Solar Nebula.
\end{itemize}

The mechanism described here has the following important properties  :
\begin{itemize}
    \item The formation of a pressure bump just \textit{inward} the snow-line as long as the icy-pebble flux is maintained and is high-enough.
    \item The accumulation of ice-free pebbles at the pressure bump, that may result in the formation of water-free planetesimal at the bump location.
    \item Pebbles accumulating at the pressure bump are inherited from beyond the snow-line due to the inward drift of icy pebbles. But they would not cross the bump, as the bump acts as a dynamical barrier, and because they are progressively incorporated into larger planetesimals.
    \item In addition to this process, beyond the snow-line, planetesimal formation can still happen at any time, following the processes described in \cite{drazkowska_2017} (like traffic jam effect or recondensation of water vapor beyond the snow-line). 
\end{itemize}

We emphasize again that this mechanism does not necessarily take place close to the outer edge of the dead-zone. Indeed there is no need for a coincidence between the location of the snow-line and the outer edge of the dead-zone (defined as $\Sigma = \Sigma_a)$. For example for having p>0.7 (a condition very favorable for the formation of the bump) the surface density at the snow-line must be larger than $1000 \Sigma_a$ (Figure \ref{Figure_equivalent_p}). For a minimum mass solar nebula, with snow-line around 2 au, the outer edge of the dead-zone would be around >10 and >100 au for $\Sigma_a=1000 kg/m^2$ and $\Sigma_a=100 kg/m^2$ respectively.

The formation of the pressure bump may lead to a strong enhancement of the dust/gas ratio in the disk midplane, necessary to trigger the streaming-instability, and thus, planetesimal formation. In Figure \ref{Figure_simu_evol}, that is our most realistic disk simulation of the present paper, we see that, at the bump,  the ratio of dust to gas surface density increases to about $\Sigma_d/\Sigma_d \simeq 0.1$. At this place the pebble stokes number is about 0.3 \citep{charnoz_2019}. So we can estimate the dust/gas ratio of volumetric densities in the midplane. Assuming the pebbles are in a vertical steady state where turbulent diffusion acts against sedimentation, the dust scale height $H_d \simeq H_g \sqrt{ \alpha /(\alpha+St)}$ \cite{drazkowska_2014}, where $\alpha$ is $\alpha_d$ here.The ratio of volumetric densities of dust and gas in the midplane is also equal to $\rho_d/\rho_g = (\Sigma_d H_g)/(\Sigma_g H_d)$. So we get $\rho_d/\rho_g=(\Sigma_d/\Sigma_g)\sqrt{(\alpha+St)/\alpha} \simeq 17$, much larger that 1. So it is very possible that planetesimal can form in the pressure bump visible in Figure \ref{Figure_simu_evol}. However, we remind the reader that other instabilities may develop, that may or may not act against concentration of dust. \cite{Hasegawa_Takeuchi_2015} investigates a viscous instability process using a new parameterization of $\alpha$ depending on the local magnetic field, and is a generalization of the parametrisation of $\alpha$ we use here. They show that at the outer edge of the dead zone, the effective $\alpha$ may become negative, leading to a viscously unstable situation where density maxima can grow without bounds because of negative effective diffusion coefficient. This instability may happen at the outer edge of the dead-zone, when the surface density becomes comparable to $\Sigma_a$. So the pressure bump formation process may take place in addition to the instability described in \cite{Hasegawa_Takeuchi_2015}. However, we do emphasize here that whereas the instability process of  \cite{Hasegawa_Takeuchi_2015} happens at the outer edge of the dead-zone, the pressure bump formation process we describe here occurs well inside the dead-zone, at the snow-line.Large scale hydrodynamical instabilities are also known to develop, like the Rosby Wave instability, in the presence of a large viscosity gradient (that may occur at the outer edge of the dead-zone) or in the presence of a strong pressure or density bump (as in our case) \citep{Regaly_2012, Mohanty_2013}. This may imply that a large scale vortex may potentially form at the pressure bump (a physics which is not captured in our simple 1D model) which could potentially concentrate pebbles, and may lead to big planetesimals formation.  In conclusion, the evolution of dust in such a bump, that could be subject to different kind of hydrodynamical instabilities should be investigated with 3D simulations in the future.

\subsection{Implications}
\subsubsection{Isolation of isotopic reservoirs}
Recent works \citep{Kruijer_2017, Nanne_2019, Kleine_2020} propose a qualitative timeline for forming and isolating the NC and CC groups, invoking a first a phase of CAIs formation close to the star, in a disk initially dominated by CC isotopic composition. It is proposed than within the first Myrs, the infalling cloud changes composition and feeds the disk's terrestrial planets region. At about 1Myr, Jupiter' formation occurs at about 5 au and split the disk in two reservoirs that follow isolated evolution, where first the iron bodies, then the chondritic bodies are formed. The innermost reservoir, the NC, progressively stops its accretionnal evolution, at about 1.5-2 Myrs due to the shut-off of incoming material because of Jupiter's barrier. Finally, at about 5 Myrs, Jupiter's reaches its final mass and scatters planetesimals inward, leading to radial mixing of the two groups. This model was recently criticized by \cite{Brasser_2020} who argues that the formation of Jupiter as late as 1Myr at 5 au would not prevent enough isolation of the CC and NC groups due to the fast delivery of pebbles in the terrestrial planets region. This delivery would lead to the isotopic contamination of the inner (NC) reservoir by material coming from the CC reservoir. It would lead also to a martian embryo too big compared to its actual mass. The authors then suggest that the isolation of the reservoirs must happen before 1Myr in order to achieve both well separated isotopic groups and slowing down of Mars' growth. In addition, recent work by \cite{Bitsch_2015, Pirani_2019,Oberg_Wordsworth_2019} suggest that Jupiter may have formed beyond 15 au and reached its final location only by the end of the disk lifetime, at around 3Myrs, well after the differentiation of the NC and CC iron bodies.

\begin{figure*}
	\includegraphics[scale=0.15]{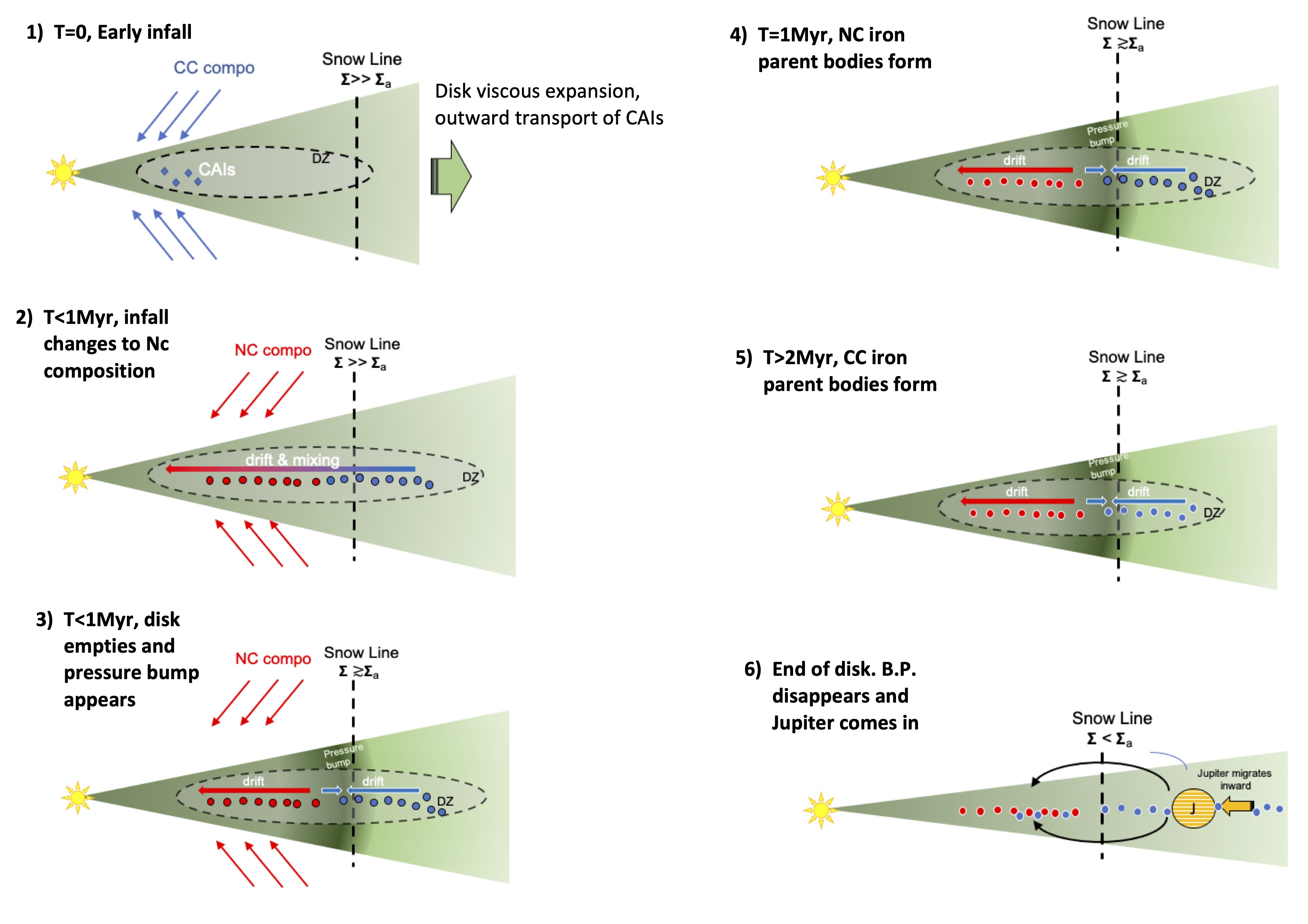}
 \caption{Schematic of the disk and NC and CC populations evolution. The green stands for the gas disk. The pressure bump is symbolised by a darker green zone, just inward the snow-line. Dashed ellipse stands for the dead-zone. Blue and red arrows show the molecular cloud gas infall. Dots with dark circle are the CC and NC bodies parent bodies. Dost with white circle stand for the irong and chondritic populations. This figure is inspired and modified from \cite{Kleine_2020}.}
  \label{disk_scheme_4panels}
\end{figure*}

The pressure bump formation mechanism presented here may be an ideal alternative to the Jupiter hypothesis. The timeline of events  proposed in \cite{Kruijer_2017}, \cite{Nanne_2019} and \cite{Kleine_2020} could be qualitatively revisited as follows (see Figure \ref{disk_scheme_4panels}). 
First, at time=0 the disk is dominated by CC isotopic composition and CAIs are formed and transported outwards during the viscous expansion of the disk \citep{pignatale_2018, Nanne_2019}. As the disk is still massive and pebbles are small (with stokes numbers < 0.01) the pressure bump cannot form. In the first few 100 Kyrs the material infalling from the cloud changes composition to NC and feeds the disk inner region \textit{inward} the snow-line. As long as the pressure bump is not established, pebble can cross the snow-line. Then as the disk empties the surface density at the snow-line drops down $< \alpha_a/\alpha_d \times \Sigma_a$ which may occur in a few 100 Kyrs only, leading to the spatial separation of the NC and CC groups.This situation lasts for a few Myr as long as flux of pebbles is maintained. During this period NC and CC groups bodies accrete separately inward and outward the snow-line respectively. Since the NC population is not fed anymore with pebbles (which are blocked at the pressure bump) it may stop growing at about 2 Myrs. At the end of the gas-disk life, when  $\Sigma <\Sigma_a$ at the snow-line, the pressure bump disappears but planetesimals are already formed and do not migrate due to gas drag. At approximately the same epoch Jupiter should come-in, migrating inward from the outer Solar System. It may scatter embryos and planetesimals leading to substantial radial mixing of the two populations.
Note that in Figure \ref{disk_scheme_4panels} the snow-line is displayed as fixed, but this is not mandatory. It is very possible that the snow-line may have migrated outward during the disk infall, then inward during the disk evolution \citep{pignatale_2018, Baillie_2019}.

\subsubsection{Forming planetesimals}
This mechanism could also explain  how dust can accumulate efficiently \textit{inward} the snow-line and form water-free planetesimals. This has challenged models in the last years \citep{drazkowska_2014, drazkowska_2017, Ida_Guillot_2016, Hyodo_2019, Hyodo2021a, Ida2021}, showing that planetesimals may form rather \textit{outward} the snow-line (due to water-vapor recondensation, traffic-jam effect and dust-gas back-reaction) and so should always be mostly water rich, which is a problem for explaining the petrology of ordinary chondrites that only show little water alteration \citep{Ida_Guillot_2016, Ida2021, Hyodo2021a}. 

In this paper, we neglected the effects of back-reaction. We note that including the back-reaction with a dead-zone structure can develop so-called the "no-drift" runaway pile-up mode of pebbles at a certain heliocentric distance, forming planetesimals, before pebbles reach the snow-line \citep{Hyodo2021b}.

Forming planetesimals in dead-zones at the snow-line is not in itself a new idea. It has been proposed in particular in \citet{Brauer_2008}. However the present work is  different because \citet{Brauer_2008} investigated the effect of ice-accumulation \textit{outward} the snow-line and showed that the increase of dust-to-gas ratio may modify locally $\alpha$ and trigger a pressure maxima. The process investigated here does not-rely on that effect, but rather on an accumulation of gas at locations of low-sound velocities like just  \textit{inward} the snow-line.

\subsection{Making pressure traps at condensation fronts}

The mechanism for forming a pressure bump detailed here may potentially work at every condensation fronts, provided the flux of heavy evaporating species is strong enough and that an evaporating front is embedded in the dead-zone. Observations of the TW-Hydra disk potentially reveals efficient trapping of CO or $N_2$ dust >10 au at their evaporation front \citep{Bosman_Banzatti_2019, McClure_2020}. However the present paper only focuses on water evaporation. The viability of such a process at the CO or $N_2$ condensation fronts must be quantified in a future study.

\section{Limitations}
Several limitations and uncertainties may limit the validity of our work. This simple model is based upon the assumption of an idealized $\alpha$ disk model with two layers (one dead, one active), which validity is still a matter of debate : see \cite{Turner_2014} for a critical review. It should be confronted, in the future, to non-ideal MHD simulations, including dust, gas and evaporation, for confirmation. 

In addition, for the pressure bump to survive it must be constantly replenished by incoming water vapor. However the water ice reservoir is limited and its emptying timescale $T_e$ is short : it is about $T_e=m M_d/(\dot M F_i/F_g)$ where $m$ is the disk's metallicity and $M_d$ is the disk mass. For $M_d= 0.02 M_{\odot}$, $F_i/F_g=0.6$ , $\dot M=10^{-8} M_{\odot}/yr$ and m=0.01 we get $T_e \simeq  17 Kyrs$. $T_e$ is much shorter than the disk lifetime, and shorter than the accretion time of the CC chondrites group, but large uncertainties exist on all parameters: the disk's mass ass well as $F_i/F_g$ and $\dot M$ that evolve with time.

A speculative solution to the short timescale of water inflow may be the following : 
\citet{Manara_2019} proposed that the disk may be continuously replenished during most of its lifetime by infalling material from the molecular cloud in order to explain the discrepancy between the measured dust content of disks in millimetric observations, and the mass distribution of exoplanets. Feeding the disk with ISM material, in stellar ratio of oxygen abundance,  would allow  maintaining the icy pebbles inflow, allowing the long-term existence of pressure maxima. This would be a major change in the paradigm of protoplanetary disk evolution, but  still controversial. This worth to consider as simulations of star formation (starting from the molecular cloud) show that infall phase could last up to a few Myr \citep{Padoan_2014}. Whereas speculative, this possibility would be worth to consider for future studies.

\subsection*{Acknowledgments}
 We wish to acknowledge the financial support  of the Region Île-de-France through the DIM-ACAV + project: “HOC -Origine de l’eau et du carbone dans le Système Solaire”, the french "Programme National de Planetologie" (PNP), the program ANR-15-CE31-0004-1 (ANR CRADLE), the IdEx Université de Paris ANR-18-IDEX-0001, and program ANR-20-CE49-0006 (ANR DISKBUILD). R.H. was supported by JSPS Kakenhi JP17J01269 and 18K13600. R.H. also acknowledges JAXA’s International Top Young program. We warmly thank S. Ida, T. Guillot and R. Brasser for useful comments. We also thank our anonymous reviewer for a thoughtful review that helped us to largely improve the quality of the paper.

 \small

\bibliographystyle{aa} 
\bibliography{references} 

\begin{appendix}

\section{A physical explanation for the viscous instability in a stratified disk}
\label{Appendix_explanation}
We give here a physical explanation for the origin of pressure bump mechanism described in the paper. We emphasize the key role played by the layered accretion structure. We track the surface density $\Sigma$ evolution of a stratified disk with two layers. The midplane layer (low  turbulence) has a surface density $\Sigma_d$ and $\alpha=\alpha_d$ and the active layer (high turbulence) has a column density $\Sigma_a$ and an $\alpha=\alpha_a$, with $\alpha_a >> \alpha_d$ . The local total surface density is $\Sigma=\Sigma_a+\Sigma_d$. There are two keys ingredients making the instability :
\begin{itemize}
    \setlength{\itemsep}{1.2\baselineskip}
    \item The active layer has a fixed surface density $\Sigma_a$
    \item there is a minimum of sound velocity somewhere (for example at the snow-line due to the release of water-vapor)
\end{itemize}

Let's imagine that at location $r_0$ there is a minimum of sound velocity so that :
\begin{equation}
    \frac{\partial C^2}{\partial r }< 0 \text{ at $r<r_0$}
\end{equation}

\begin{equation}
    \frac{\partial C^2}{\partial r }> 0 \text{ at $r>r_0$}
\end{equation}

How is the mass is transported in the disk at $r_0$? The total mass flux  F (i.e the local accretion rate) can be split in two contributions: $F=F_d+F_a$ ($F_d$= mass flux in the DZ, and $F_a$= mass flux in the active layer). It is useful to rewrite the surface density equation in terms of divergence of mass flux: The surface density evolution is :

\begin{equation}
\frac{\partial \Sigma}{\partial t} = \frac{3}{r}\frac{\partial}{\partial r}\left( r^{1/2} \frac{\partial}{\partial r}(\nu \Sigma r^{1/2})\right) 
\end{equation}
Introducing F=local mass flux, the variation of density is always opposite to the divergence of the material flux:

\begin{equation}
\frac{\partial \Sigma}{\partial t} = \frac{-1}{2\pi r}\frac{\partial}{\partial r}\left( F \right) 
\end{equation}

with :
\begin{equation}
F =- 6 \pi r^{1/2} \frac{\partial}{\partial r}\left( \frac{\alpha C^2 }{\Omega} \Sigma r^{1/2}  \right) 
\end{equation}
We write $\Omega=B r^{-3/2}$. Considering the dead-zone and the active layers as two layers with constant $\alpha$  and developing the derivative we have:

\begin{equation}
F_d =\frac{- 6 \pi \alpha_d r^{1/2}}{B}  \left( 2 r C^2 \Sigma_d + r^2 C^2 \frac{\partial \Sigma_d}{\partial r}+r^2\Sigma_d \frac{\partial C^2}{\partial r}  \right) 
\label{Eq_Fd}
\end{equation}
This is the flux  in the DZ. We see that if there is an overdensity of $\Sigma_d$ in the dead-zone(the second term), if strong enough ,  $F_d$ will be opposite to the density perturbation, and the viscous evolution will erase the density perturbation. This is the classical view of viscous diffusion where the material flux is opposite to density variations.
Now If we do the same calculation for the active layer, and keeping in mind that the surface density of the active layer is constant  $=\Sigma_a$ we get for the radial flux in the active layer:

\begin{equation}
F_a =\frac{- 6 \pi \alpha_a r^{1/2}}{B}  \left( 2 r C^2 \Sigma_a+ r^2\Sigma_a \frac{\partial C^2}{\partial r}  \right) 
\end{equation}

 We see that if in the active layer  $\partial C^2/ \partial r$ is strong enough (in absolute value), the flux is \textit{always} directed toward minima of sound velocity ($F_a$ is positive when $C^2$ drops). So more material will be brought toward sound-velocity minima. But since there cannot be no more that $\Sigma_a$ of column density in the active layer, the incoming material will be transfered down to the dead-zone (that can accept it without limit), so $\Sigma_d$ will increase. In other terms, the active layer will feed the dead-zone with new material at places of low-sound velocity (at $r=r_0$ with our notations). The net result is that $\Sigma$ will increase at $r_0$. 

Checking Equation \ref{Eq_Fd} we see that when $\Sigma_d$ will become very big at $r_0$, so $\partial \Sigma_d/ \partial r$ term will become very strong at some points and will compensate for other terms and, eventually, $F_d$ will be directed away from maxima of surface density. So $\Sigma_d$ cannot diverge to infinity. Rather, there will be a transient phase with a strong increase of surface density and creation of a pressure bump as described above.

Of course this is a simplified 2 layers model. But the key idea is that the column density of the active layer is almost constant it does not "feel" density maxima in the DZ. So as the gas accretes, the material will be transporteds and stored in the dead-zone with low viscosity.

In total this creates a situation where the vertically averaged diffusion coefficient $\nu$ decreases with increasing $\Sigma$, which is typical of a viscous instability process, favoring an enhancement of high surface-density regions.

\begin{figure} 
\centering
	\includegraphics[width=0.4\textwidth]{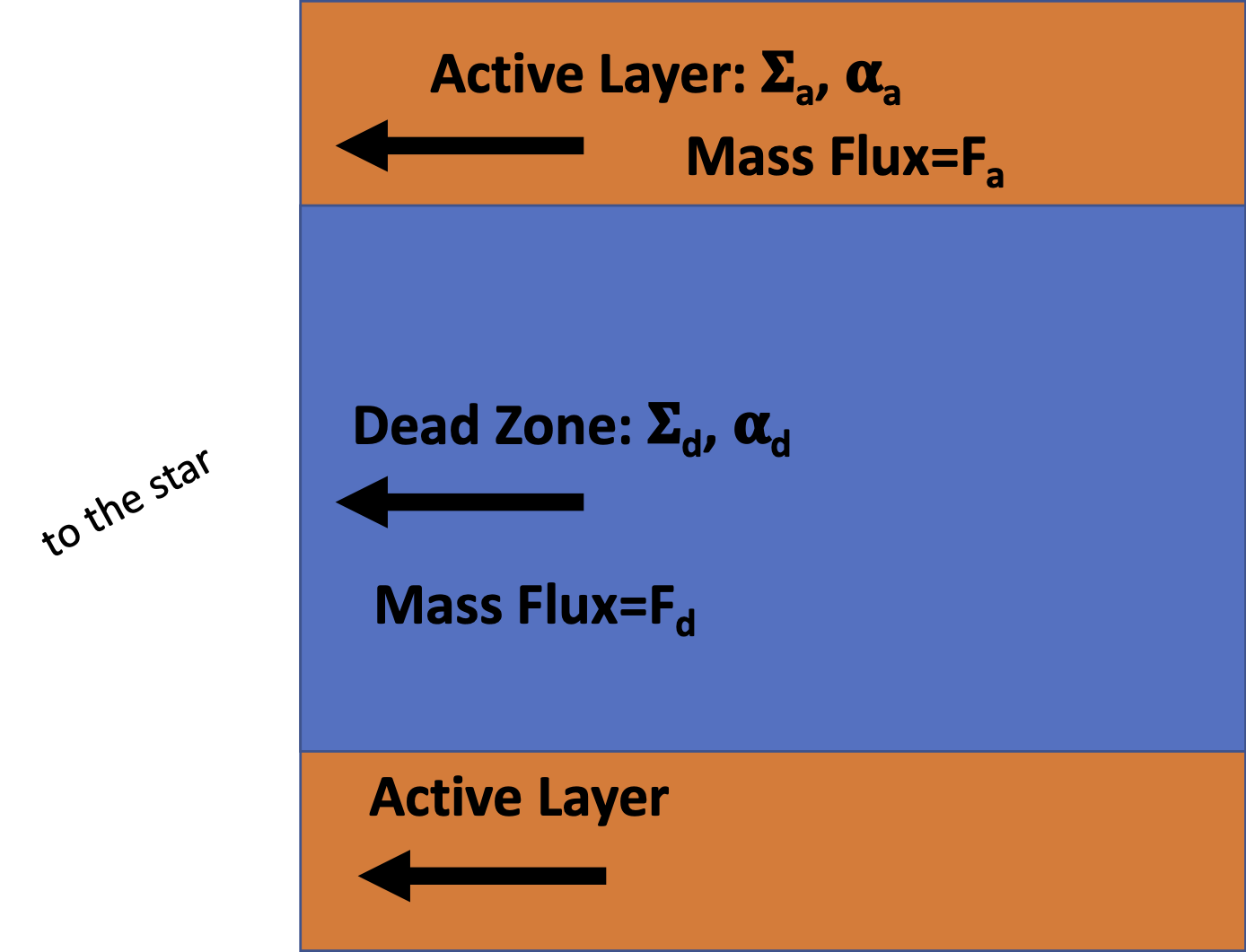}
\caption{Sketch of the two layers in the disk with their respective mass fluxes.}\label{sketch_2_layers}
\end{figure}
\end{appendix}
\end{document}